\documentclass[12pt]{article}
\usepackage{latexsym}
\usepackage{amssymb,cite}
\usepackage{amsmath,epsfig}
\usepackage{rotating}
\usepackage{graphics,color}
\usepackage{a4}
\newcommand{\R}{\mathbb{R}}

\newcommand{\cC}{{\cal C}}
\newcommand{\cD}{{\cal D}}

\newcommand{\cM}{{\cal M}} 
\newcommand{\cO}{{\cal O}}

\def\be#1\ee{\begin{equation}#1\end{equation}}

\newcommand{\bra}{\langle}
\newcommand{\ket}{\rangle}

\def\be#1\ee{\begin{equation}#1\end{equation}}
\def\bea#1\eea{\begin{align}#1\end{align}}

\begin{document}

\title{\bf\large 
Complex Langevin: Etiology and Diagnostics of its Main Problem
}

\author {
\addtocounter{footnote}{2}
 Gert Aarts$^a$\thanks{email: g.aarts@swan.ac.uk} \,\,\,\,
 Frank A.~James$^a$\thanks{email: pyfj@swan.ac.uk} \,\,\,\,
\addtocounter{footnote}{1}
 Erhard Seiler$^b$\thanks{email: ehs@mppmu.mpg.de}
\\ and
 Ion-Olimpiu Stamatescu$^c$\thanks{email:
I.O.Stamatescu@thphys.uni-heidelberg.de} \\
\mbox{} \\
 {$^a$\em\normalsize Department of Physics, Swansea University} \\
 {\em\normalsize Swansea, United Kingdom} \\
 \mbox{} \\
 $^b${\em\normalsize Max-Planck-Institut f\"ur Physik   
(Werner-Heisenberg-Institut)} \\
 {\em\normalsize M{\"u}nchen, Germany} \\
 \mbox{} \\
 $^c${\em\normalsize Institut f\"ur Theoretische Physik, Universit\"at
Heidelberg and FEST} \\
 {\em\normalsize Heidelberg, Germany} \\
}

\date{} 
\maketitle 

\begin{abstract} \noindent The complex Langevin method is a leading 
candidate for solving the so-called sign problem occurring in various 
physical situations. Its most vexing problem is that in some cases it 
produces `convergence to the wrong limit'. In the first part of the paper 
we go through the formal justification of the method, identify points at 
which it may fail and identify a necessary and sufficient criterion for 
correctness. This criterion would, however, require checking infinitely 
many identities, and therefore is somewhat academic. We propose instead a 
truncation to the check of a few identities; this still gives a necessary 
criterion, but a priori it is not clear whether it remains sufficient. In 
the second part we carry out a detailed study of two toy models: first we 
identify the reasons why in some cases the method fails, second we test 
the efficiency of the truncated criterion and find that it works perfectly 
at least in the toy models studied. 
\end{abstract} 
\maketitle

\section{Introduction} 
\label{secI}

The sign problems arising in simulations of various systems, in particular 
in QCD with finite chemical potential \cite{pdf}, are in principle solved 
by using the complex Langevin equation (CLE). This method, after being 
proposed in the early 1980s by Klauder \cite{klauder1} and Parisi 
\cite{parisi}, enjoyed a certain limited popularity (see for instance 
\cite{Karsch:1985cb,Damgaard:1987rr}) and has in more recent years been 
revived with some success \cite{Berges:2005yt, Berges:2006xc, 
Berges:2007nr, Aarts:2008rr,Aarts:2008wh, ass1, Aarts:2009hn, 
Aarts:2010gr, Aarts:2010aq}. Unfortunately already in the beginning 
problems were encountered. The first problem, instability of the 
simulations (runaways) can be dealt with by introducing an adaptive step 
size, as shown in \cite{ass1}. More vexing is the second problem: 
convergence to a wrong limit \cite{Ambjorn:1985iw, klauder2, 
linhirsch,Ambjorn:1986fz}. It is this problem which we wish to address in 
this paper.

A formal argument for correctness of the CLE was presented in a previous 
paper \cite{ass2}. It proceeded by comparing two time evolutions: the 
first one of a complex measure not allowing a probabilistic interpretation
-- the origin of the sign problem -- , the other one of a positive measure 
on a complexified space, allowing a probabilistic interpretation and hence 
suitable for simulation. The main point was that, ignoring certain 
subtleties, these two time evolutions led to identical evolutions     
for the expectation values of holomorphic observables. This implied of 
course also that the long-time limits (assuming their existence) agreed, 
corresponding to the desired equilibrium expectation values.

In \cite{ass2} we already identified some difficulties with those formal 
arguments. Some of them were of a slightly academic nature, namely the 
mathematically sticky problem of the existence of those evolutions and 
their convergence properties. Taking a pragmatic attitude, these problems 
are answered by performing simulations; in a large set of examples the 
answer is positive. The remaining problem is much more insidiuous: it may 
(and sometimes it does) happen that the results of a simulation are well 
converged and look perfectly fine, but turn out to be wrong when compared 
with known results. Of course in most interesting cases the results are 
not known and one would like to have a test to decide if one should trust 
the outcome of a simulation or not.

To deal with the disease of `convergence to the wrong limit' it helps to 
have a deeper understanding of its causes -- the etiology -- before 
developing ways to diagnose it. The possible causes of the failure of the 
formal arguments are: insufficient falloff of the probability distribution 
in the imaginary directions and too strong growth of the (time-evolved) 
observables in the imaginary directions. They can invalidate the 
integrations by parts that are necessary to show agreement of the two time 
evolutions mentioned above. Section \ref{secII} revisits the formal 
argument and deduces a crucial identity (depending on the observable 
chosen) that has to be fulfilled if the two time evolutions are to agree.

Section \ref{secIII} considers the long-time limits of the two time 
evolutions. Our set of identities (one for each observable)  leads in the 
long time limit to a set of simpler ones which turn out to be closely 
related to the Schwinger-Dyson equations. We show that in principle the 
complete set of these identities, together with a certain bound, is 
necessary and sufficient to establish correctness, provided some mild 
technical conditions are fulfilled. For practical purposes, however, the 
infinite set of identities has to be truncated to a finite (actually 
small) set; a proof of their sufficiency is thus no longer possible.

We then study these issues in detail in two toy models. The first one is a 
one-link version of lattice U(1) gauge theory, already studied in 
\cite{Aarts:2008rr,ass2}; the second one, which was first studied   
by Guralnik and Pehlevan \cite{gp1}, is a polynomial model with 
purely imaginary action, which is a toy version of the real time Feynman 
path integral. In this investigation we use noise both in the real and the 
imaginary directions, even though in principle real noise would be 
sufficient. The reason for this is twofold: it allows to study how and why 
the formal arguments can fail, and it gives us the possibility to approach 
the problem also in the dual way by solving the Fokker-Planck equation.

In Section IV we study the identity necessary for the agreement of the two 
time evolutions for the U(1) one-link model. In order to be able to do 
this with sufficient precision, we introduce a device that at first seems 
rather ad hoc: we introduce a periodic cutoff for the imaginary part of 
the field. But we can take advantage of the fact that a positive measure 
on the complexification of the original field space, required only to 
produce the right expectations for {\it holomorphic} observables, is not 
unique, so it is conceivable that a measure with cutoff gives correct 
results. With a cutoff we can compute the time evolution of the 
probability measure with high precision using the Fokker-Planck equation 
(FPE). The time evolution of the observables (averaged over the noise), is 
also needed; the way we define it here it does not depend on the cutoff.  
Of course the cutoff invalidates the formal argument, and as expected, we 
find that the two evolutions in general do not agree. But surprisingly it 
turns out that the cutoff can be tuned to a value that restores agreement.

But how can the formal argument fail when there is no cutoff?  This is 
explained by looking at the growth of the (noise averaged) observables in 
imaginary direction, evolved for a finite amount of time. It turns out 
that these averages grow like an exponential of an exponential, a growth 
that cannot be compensated by the decay of the probability measure; so the 
formal argument becomes inapplicable.

In Section V we investigate for both toy models the falloff of the 
equilibrium distribution in imaginary direction; again it is found that in 
the presence of complex noise the falloff is insufficient for the 
derivation of the SDE identities. This corroborates the indications 
presented in \cite{ass2}. On the other hand, for only real noise the 
distributions show much stronger falloff (actually they are concentrated 
on a line in the U(1) one-link model), which is sufficient for the 
derivation of the SDE identities.

Finally in Section VI we use a truncated form of our SDE criterion as a 
test of correctness of the equilibrium measures in both our toy models; it 
turns out that the test is surprisingly strong. To put it in terms of 
medical statistics: the test has perfect {\it specificity} (100\%), i.e. 
when the simulation is correct, it is always fulfilled; this is a general 
mathematical fact. But the pleasant surprise is its very strong {\it 
sensitivity}, meaning that in the cases studied, when it is fulfilled, the 
results, as far as checked, are correct. 

Here again we introduce a periodic cutoff for the imaginary part of the 
field variable. As in the finite time situation this can in general not be 
expected to work, because it destroys the formal argument for correctness, 
but again it turns out that in the two toy models studied here, the cutoff 
can be tuned to produce correct results. Requesting fulfillment of a few 
of the identities mentioned above is then used as a test for correctness: 
surprisingly we find not only that this test can be fulfilled by tuning 
the cutoff, but that in this case we obtain the correct expectation 
values. The same situation arises for real noise: the simulations pass the 
test and produce the right values. However, in lattice models the 
situation is not so simple and just suppressing imaginary noise is not 
always sufficient. This has already been found in the XY model 
\cite{Aarts:2010aq}; in a separate paper \cite{ajss} that model will be
analyzed further from the point of view developed in this article.

Finally in Sec.~\ref{secVII} we draw some conclusions and present an 
outlook on work in progress. 

\section{The formal arguments revisited}
\label{secII}

We briefly go through the arguments presented in \cite{ass2}, 
concentrating on models in which the fields take values in flat manifolds 
${\cM_r}=\R^n$ or ${\cM_r}=T^n$, where $T^n$ is the $n$ dimensional 
torus $(S^1)^n$ with coordinates $(x_1,\ldots,x_n)$.

The complex measure $\exp(-S)dx$, with $S$ a holomorphic function on a 
real manifold $\cM$, is replaced by a positive measure $Pdxdy$ on the 
complexification $\cM_c$ of $\cM$, which is the equilibrium measure of the 
complex Langevin process on $\cM_c$; the hope is that expectation values 
of {\it entire holomorphic observables} $\cO$ agree with those obtained 
using the complex measure $\exp(-S) dx$.

The complex Langevin equation (CLE) on $\cM_c$ is
\begin{align}
dx=&K_x dt +\sqrt{N_R}dw_R,\notag\\
dy=&K_y dt +\sqrt{N_I}dw_I,
\label{cle2} 
\end{align}
where $dw_R$ and $dw_I$ are independent Wiener processes, $N_I\ge 0$ and 
$N_R=N_I+1$. In the case $N_I>0$ we speak of complex noise. The drift is 
given by
\begin{align}
K_x=& -{\rm Re} \nabla_x S(x+iy), \notag \\
K_y=& -{\rm Im} \nabla_x S(x+iy).
\end{align}

By It\^o calculus, if $f$ is a twice differentiable function on 
$\cM_c$ and 
\be
z(t)=x(t)+iy(t)
\ee 
is a solution of the complex Langevin equation (\ref{cle2}), we have 
\be
\label{ito}
\frac{d}{dt}\left\bra f(x(t),y(t))\right\ket = \left\bra L 
f(x(t),y(t))\right\ket,
\ee
where $L$ is the Langevin operator 
\be
\label{eq:LO}
L=\left[N_R\nabla_x+K_x\right] \nabla_x  
+ \left[N_I \nabla_y+K_y\right] \nabla_y,
\ee
and $\bra f \ket$ denotes the noise average of $f$ corresponding to the 
stochastic process described by Eq.~(\ref{cle2}). In the standard way 
Eq.~(\ref{cle2}) leads to its dual Fokker-Planck equation (FPE) for the 
evolution of the probability density $P(x,y;t)$,
 \be
\label{realFPE}
\frac{\partial}{\partial t} P(x,y;t)= L^T P(x,y;t),
\ee
with 
\be
L^T=\nabla_x\left[N_R\nabla_x-K_x\right]+
\nabla_y\left[N_I\nabla_y- K_y\right].
\ee
$L^T$ is the formal adjoint (transpose) of $L$ with respect to the 
bilinear (not hermitian) pairing 
\be
\bra P, f\ket= \int f(x,y) P(x,y) dxdy, 
\ee
i.e.,
\be
\bra P, Lf\ket= \bra L^T P, f\ket.
\ee
To understand the relations between the real and the complex measures one 
has to consider the evolution of a complex density $\rho(x)$ on $\cM$ 
under the following complex FPE 
\be  
\label{complexFPE}
\frac{\partial}{\partial t} \rho(x;t)= L_0^T \rho(x;t),
\ee
where now the complex Fokker-Planck operator $L_0^T$ is  
\be
\label{fpc0}
L_0^T =  \nabla_x \left[\nabla_x+(\nabla_x S(x))\right].
\ee
We will also use a slight generalization: For any $y_0\in \cM$ we consider
the complex Fokker-Planck operator $ L_{y_0}^T$ given by
\be
\label{fpc1}
L_{y_0}^T=\nabla_x \left[\nabla_x+(\nabla_x S(x+iy_0))\right].
\ee
$L_{y_0}^T$ is the formal adjoint of
\be
L_{y_0}=  \left[\nabla_x-( \nabla_x S(x+iy_0))\right]\nabla_x.
\ee
The complex density
\be 
\label{rhostat}
\rho(x;\infty)\propto \exp\left[-S(x)\right] 
\ee
is a stationary solution of Eq.~(\ref{complexFPE}), which is expected to 
be unique. Numerical studies (where feasible) of Eq.~(\ref{complexFPE}) 
confirm this; in fact the convergence to the limit Eq.~(\ref{rhostat}) 
seems to be exponentially fast.

We have to make a few technical remarks about the space of observables we 
choose: all observables have to be entire holomorphic functions; we will 
furthermore require that their restrictions to the real submanifold 
$\cM_r$ span a large enough space $\cD$:

(1) if $\cM_r=T^n$, $\cD$ should be a dense subset of $\cC(\cM_r)$, the 
set of all continuous functions on $\cM$ equipped with the norm 
$||\cO||\equiv\sup_x|\cO(x)|$; a good choice is the space of finite linear 
combinations of exponentials.

(2) if $\cM_r=\R^n$ and the action $S$ has a real part that grows at least 
like $|x|$ as $|x|\to\infty$, the functions in $\cO\in \cD$ should be 
bounded polynomially and dense in the Banach space defined by the norm 
$||\cO||\equiv\sup_x \exp(-|x|) |\cO(x)|$; a natural choice for $\cD$ is 
the space of polynomials.
 
(3) if $\cM_r=\R^n$ and the action is purely imaginary, one has to find a 
submanifold $\cM'_r\subset \cM_c$ which is a suitable deformation of 
$\cM_r$ into the complex domain, such that the integral of $\exp(-S)$ 
converges and $\cM'_r$ can still be parameterized by $x\in\R^n$. The 
conditions on the obervables, expressed in this parameterization are then 
as in (2). In a slight abuse of language, we still refer to $\cM'_r$ as 
the `real submanifold'. Again polynomials are a natural choice for the 
space of observables.

We set
\be
\label{eq:OP}
\bra \cO\ket_{P(t)}\equiv \frac{\int O(x+iy) P(x,y;t) dxdy}
{\int  P(x,y;t) dxdy}
\ee
and
\be
\bra \cO\ket_{\rho(t)}\equiv \frac{\int O(x) \rho(x;t) dx}
{\int\rho(x;t) dx} .
\ee  
What one would like to show is that
\be
\label{correctness}
\bra \cO\ket_{P(t)}=\bra \cO\ket_{\rho(t)},
\ee
if the initial conditions agree,
\be
\bra \cO\ket_{P(0)}=\bra \cO\ket_{\rho(0)},
\ee
which is assured provided 
\be 
\label{init}
P(x,y;0)=\rho(x;0)\delta(y-y_0)\,.
\ee 
One expects that in the limit $t\to\infty$ the dependence on the initial 
condition disappears by ergodicity.\footnote{In \cite{Aarts:2010gr} 
dependence on initial conditions was found to be due to peculiar features 
of the classical flow pattern, leading to degenerate equilibrium 
distributions.}

To establish a connection between the `expectation values' with respect to 
$\rho$ and $P$ for a suitable class of observables, one moves the time 
evolution from the densities to the observables and makes use of the 
Cauchy-Riemann (CR) equations. Formally, i.e.~without worrying about 
boundary terms and existence questions, this works as follows: first we 
use the fact that we want to apply the complex operators $L_{y_0}$ only 
to functions that have analytic continuations to all of $\cM_c$. On those 
analytic continuations we may act with the Langevin operator
\be
\tilde L \equiv \left[\nabla_z-(\nabla_z S(z))\right]  \nabla_z,
\ee
whose action on holomorphic functions agrees with that of $L$, since on 
such functions $\nabla_y=i\nabla_x$ and $\Delta_x =-\Delta_y$ so that the 
difference $L-\tilde L$ vanishes.

We now use $\tilde L$ to evolve the observables according to the equation
\be
\label{obsevol}
\partial_t \cO(z;t)= \tilde L \cO(z;t)\quad (t\ge 0)
\ee
with the initial condition $\cO(z;0)=\cO(z)$, which is formally solved by
\be
\label{obssol}
\cO(z;t) =  \exp[t \tilde L] \cO(z).
\ee
In Eqs.~(\ref{obsevol}, \ref{obssol}), because of the CR equations, the 
tilde may be dropped, and we will do so now. So we also have  
\be
\label{obssol2}
\cO(z;t) = \exp[t L] \cO(z).
\ee
In \cite{ass2} it was shown that $\cO(z;t)$ is holomorphic if $\cO(z;0)$ 
is. The evolution can therefore also be obtained equivalently by solving 
\be
\label{obsevol2}
\partial_t \cO(x+iy_0;t)=  L_{y_0} \cO(x+iy_0;t)\quad (t\ge 0)
\ee
and subsequent analytic continuation.

The crucial object to consider is, for $0\le \tau\le t$, 
\be
F(t,\tau)\equiv \int P(x,y;t-\tau) \cO(x+iy;\tau)dxdy,
\label{fttau}
\ee
which interpolates between the $\rho$ and the $P$ expectations:
\be
F(t,0)= \bra \cO\ket_{P(t)}, \;\;\;\; F(t,t)= \bra \cO \ket_{\rho(t)}.
\ee
The first equality is obvious, while the second one can be seen as 
follows, using Eqs.~(\ref{init}, \ref{obssol2}), 
\begin{align} 
F(t,t)=&\int P(x,y;0) \left(e^{t L}\cO\right)(x+iy;0)dxdy\notag\\=&
\int \rho(x;0) \left(e^{tL_0} \cO\right)(x;0)dx\notag\\=&
\int \cO(x;0)\left(e^{tL_0^T}\rho\right)(x;0)dx\notag\\ = &
\bra \cO \ket_{\rho(t)},
\end{align}
where it is only necessary to assume that integration by parts in 
$x$ does not produce any boundary terms. 

The desired result Eq.~(\ref{correctness}) would follow if $F(t,\tau)$ 
were independent of $\tau$. To check this, we take the $\tau$ derivative:
\begin{align}
\label{interpol}
\frac{\partial}{\partial \tau} F(t,\tau) = &
-\int \left(L^T P(x,y;t-\tau)\right)\cO(x+iy;\tau)dxdy\notag\\
& + \int  P(x,y;t-\tau) L\cO(x+iy;\tau) dxdy.
\end{align}
Integration by parts, if applicable without boundary term at infinity, 
then shows that the two terms cancel, hence $\frac{\partial}{\partial 
\tau} F(t,\tau)=0$  and thus proves Eq.~(\ref{correctness}), irrespective 
of $N_I$. 

So here we have found a place where the formal argument may fail: if the 
decay of the product 
\be
P(x,y;t-\tau)\cO(x+iy;\tau)
\ee
and its derivatives is insufficient for integration by parts without 
boundary terms.

If (\ref{interpol}) vanishes and furthermore
\be
\label{conv}
\lim_{t\to \infty} \bra \cO\ket_{\rho(t)} = \bra \cO \ket_{\rho(\infty)},
\ee
with $\rho(\infty)$ given by Eq.~(\ref{rhostat}), one can conclude 
that the expectation values of the Langevin process relax to the 
desired values. Eq.~(\ref{conv}) requires that the spectrum of $L^T_{y_0}$ 
lies in a half plane ${\rm Re}\,z\le 0$ and $0$ is a nondegenerate 
eigenvalue. (Actually, convergence of $P(x,y;t)$ is more than what is 
really needed, because the measure will only be tested against holomorphic 
observables.)

The numerical evidence in practically all cases points to the existence 
of a unique stationary probability density $P(x,y;\infty)$. More detailed  
information about this will be given below.

In \cite{ass2} three questions were raised. The first one concerned the 
exponentiation of the operators $L, \tilde L$ and their transposes, or in 
other words whether they are generators of semigroups on some suitable 
space of functions. Even though we have not found a general mathematical 
answer to this question, numerics indicate that it is affirmative in all 
cases considered; for $L_{y_0}$ in our first toy model a proof will be 
given in Sec.~\ref{secIV}. Likewise it is not known whether the spectra of 
$L, L_{y_0}$ are contained in the left half plane and if $0$ is a 
nondegenerate eigenvalue, but the numerics again strongly indicate an 
affirmative answer.

So the main remaining question concerns the integrations by parts without 
boundary terms, which underlie the shifting of the time evolution from the 
measure to the observables and back; actually what is really needed is the 
ensuing $\tau$ independence of $F(t,\tau)$, defined in Eq.~(\ref{fttau}).

A crucial role for the correctness of CLE simulations is therefore played 
by the vanishing of (\ref{interpol}). Whether this holds or not will be 
studied in detail for one of our toy models in Section \ref{secVI}. 

\section{A criterion for correctness}
\label{secIII}

As explained in the previous section, $F(t,\tau)$  has to be independent 
of $\tau$ for all times $t$, i.~e.
\be
\frac{\partial}{\partial\tau}F(t,\tau)=0\,.
\label{critid}
\ee
Below in Section IV it will be seen that for the U(1) one-link model the 
$\tau$ derivative is largest at $\tau=0$. This motivates to try the 
superficially weaker condition
\be
\lim_{t\to\infty}\frac{d}{d\tau}F(t,\tau)\biggr|_{\tau=0}=0\,.
\label{crit0}
\ee
We will see later that this condition is in fact still sufficient for 
correctness, modulo some technical conditions, if it holds for a 
sufficiently large set of observables.

If we now look again at Eq.~(\ref{interpol}), we realize that for the
equilibrium measure (always assuming it exists) $L^T P(x,y;\infty)=0$ and 
hence the first term on the right hand side vanishes. The criterion 
(\ref{crit0}) thus turns into
\be
E_{\cO}\equiv\int P(x,y;\infty) \tilde L \cO(x+iy;0)dx dy
=\bra \tilde L\cO\ket=0\,,
\label{crit}
\ee
where we used the fact that on $\cO$ $L$ and $\tilde L$ can be used 
interchangeably. This would of course also follow from the equilibrium 
condition $L^T P(x,y;\infty)=0$ on $\cM_c$, if the decay of $P$ at large 
$y$ is sufficient to allow integration by parts on $\cM_c$ without 
boundary term. Eq.~(\ref{crit}) is a fairly simple condition that is 
rather easy to check for a given observable. But it has to be satisfied 
for `all' observables i.e.~for a basis (in a suitable sense) of our 
chosen space $\cD$, so it represents really an infinite tower of 
identities.

It may be worth noting that the collection of identities (\ref{crit}),
applied to all observables, is closely related to the Schwinger-Dyson 
equations (SDE). We show this for the simple case of a scalar theory on a 
lattice with fields denoted by $\phi_i$ : the SDEs are well-known to 
arise from the relation
\be
\left\bra \frac{\partial f}{\partial \phi_i} \right\ket=\left\bra f
\frac{\partial S}{\partial \phi_i}\right\ket\,
\label{SDE}
\ee
for `any' function $f$ of the fields (in most applications the observables
are chosen to be exponentials $\exp(\sum_i \phi_i j_i)$). Our Langevin
criterion $\bra\tilde L{\cO}\ket$ on the other hand reads
\be
\sum_i \left\bra \frac{\partial^2 \cO} {\partial \phi_i^2}\right\ket=
\sum_i \left\bra \frac{\partial \cO}{\partial \phi_i}
\frac{\partial S} {\partial \phi_i}\right\ket\,.
\label{LE}
\ee
It is quite obvious that Eq.~(\ref{SDE}) implies Eq.~(\ref{LE}): we only
have to set in Eq.~(\ref{SDE}) $f=\partial_i \cO$. The converse is also
easy: we only have to find a set of observables $\cO_j$ satisfying
\be
\sum_i\frac{\partial^2\cO_j}{\partial\phi_i^2}=\partial_j f\,;
\ee
this involves just inversion of the (functional) Laplace operator, which  
is always possible here, because the only zero modes are constants.
   
We proceed to show that in principle the identities for a sufficiently 
large (countably infinite) set of obervables are also sufficient to assure 
correctness, provided a certain bound is satisfied. Let us now assume 
that we have, by whatever method, obtained a measure $Q$ on $\cM_c$ that 
allows integration of all $\cO\in\cD$ and furthermore satisfies a bound
\be
|\bra Q, \cO\ket |\le C ||\cO||\,,
\label{bound}
\ee
where $C$ is some constant and the norm is the one discussed in Section
II (recall that this norm only involved the values of $\cO$ on $\cM_r$).  
We claim that modulo certain technical conditions the fulfillment of
Eq.~(\ref{crit}) for a basis of $\cD$, 
\be
\bra Q,\tilde L \cO\ket = \int Q(x,y)\tilde L\cO(x+iy)\, dx dy = 
0\,, 
\label{crit'}
\ee
implies that the $Q$ expectations are correct, i.e.
\be
\bra Q,\cO\ket = \int Q(x,y)\cO(x+iy) dx dy
=\frac{1}{Z} \int_{\cM_r} \cO(x) e^{-S(x)} dx\,.
\ee
The argument uses the fact that the values of $\cO$ on $\cM_r$ 
already determine the values on $\cM_c$. So $\bra Q,\cO\ket $ can be 
viewed as a linear functional on the space $\cD$ considered as functions 
on $\cM_r$, which is assumed to be dense in $\cC(\cM_r)$. Because of the 
bound Eq.~(\ref{bound}) this functional has a unique extension to a linear 
functional on all of $\cC(\cM_r)$. By a standard theorem of analysis -- 
the Riesz-Markov theorem (see for instance \cite{RS}) -- this linear 
functional is therefore given by a complex measure $\sigma_Q dx$ on $\cM$, 
i.e.~we can write
\be
\bra Q,\cO\ket =\int_{\cM_r} \cO(x) \sigma_Q(x) dx\,,
\ee
where $\sigma_Q$ is allowed to contain $\delta$ functions. Since $\cO$ was 
any observable, we may replace it by $\tilde L \cO$; we then have
\be
\bra Q,\tilde L \cO\ket= \int_{\cM_r} (L \cO)(x) \sigma_Q(x) dx = 0\,,
\ee
which is equivalent to
\be
\int_{\cM_r} \cO(x) (L_0^T\sigma_Q)(x)  dx = 0\,,
\ee
using only integration by parts on $\cM_r$, which in general 
unproblematic. Since this holds for all $\cO$ in the dense set $\cD$, we 
conclude
\be
L^T_0\sigma_Q = 0\,.
\ee
To deduce from this that $\sigma_Q=\exp(-S)/Z$ we only need that $0$
is a nondegenerate eigenvalue of $L^T$, an assumption we had to make
anyway in Section II. In concrete models this needs checking, of course.

If, on the other hand, we find $E_{\cO} \neq 0$ for some observable $\cO$, 
this means that our simulation is not correct. Since by formal integration 
by parts on $\cM_c$ the equilibrium condition $L^T P(x,y;\infty)=0$ would 
imply $E_{\cO} = 0$, we can see only one possible reason for $E_{\cO} \neq 
0$, namely insufficient falloff of the equilibrium measure in imaginary 
direction.

This whole discussion is a bit superficial, as far as the functional
analysis is concerned, but it is not worth going into more detail here,
since it is quite academic anyway. In practice it will be difficult to
check the bound Eq.~(\ref{bound}) (see however Section VI) and impossible 
to check the criterion for a full basis of observables; we are reduced to 
checking it for a few. So the sensitivity of the resulting test needs to 
be checked experimentally.
   
It is well known that the SDE's have spurious, unphysical solutions (see 
for instance \cite{esdoc}, \cite{Berges:2006xc} or \cite{gp2}). This
should be obvious from the fact that they are equivalent to a (functional) 
differential equation which requires at least some kind of boundary 
conditions for definiteness and also from the fact that they are {\it 
recursive} relation that can always be fulfilled by fixing the low 
moments/modes in an arbitrary way. So it has to be checked whether 
requiring the criterion Eq.~(\ref{crit}) in fact selects the correct 
expectation values. The bound Eq.~(\ref{bound}) will in general be 
sufficient for this. In Section \ref{secVI} we will see how this works in 
the U(1) one-link model.

\section{Detailed study of $F(t,\tau)$ for the U(1) one-link model}
\label{secIV}

\subsection{Numerical study}

The U(1) one-link model was introduced in \cite{Aarts:2008rr} and studied 
further in \cite{ass2}. At lowest order in the hopping expansion it is 
defined by the action
\be
S=-\beta \cos z -\kappa \cos(z-i\mu)= - a\cos(z-ic)\,,
\ee
with  
\be
a=\sqrt{(\beta+\kappa e^\mu)(\beta+\kappa e^{-\mu)}}\,,
\ee
\be
c=\frac{1}{2}\ln\frac{\beta+\kappa e^\mu}{\beta+\kappa e^{-\mu}}\,,
\ee
leading to the drift
\begin{align}
K_x&=-{\rm Re}\, S'= -a\sin x\cosh(y-c), \\
K_y&=-{\rm Im}\, S'= -a\cos x \sinh(y-c).
\end{align}
It is easy to see by shifting an integration contour that no essential 
generality is lost if we set $c=0$. So in the sequel we will make this 
choice. We will also set $a=1$. 

A natural choice of a basis for the space of observables are the 
exponentials $e^{ikz}$. Here we study in detail the question whether 
the quantity $F(\tau,t)$, see Eq.~(\ref{fttau}), is indeed independent of 
$\tau$, as required for correctness. We use both CLE and FPE for this 
analysis; since the former yields ambiguous results for $k>1$ if $N_I>0$,
whereas the latter requires $N_I>0$ for stability (see below), we
are forced to introduce a field cutoff in this analysis. We are aware of
the fact that such a cutoff destroys the formal argument for correctness,
but using the nonuniqueness of the positive measure on $\cM_c$ there is 
still a chance to get correct results with such a measure; we will
check whether it is possible by tuning the cutoff.

We introduce the cutoff in the simplest possible way by imposing periodic 
boundary conditions in field space. In our U(1) one-link model we have 
periodic b.c. in the $x$ direction by definition, so we only have to cut 
off the imaginary part; we denote the value of the cutoff by $Y$, such 
that $-Y\le y\le Y$. Periodizing the observable of course violates the 
Cauchy-Riemann (CR) equations at the `seam', while the drift becomes 
discontinuous across the `seam' making the interpretation of the FPE also 
difficult. But quite independent of those issues, if Eq.~(\ref{critid}) 
holds the equality Eq.~(\ref{correctness}) follows and thus the 
correctness of the CLE method is assured.

In any case we will see that our rather naive cutoff procedure seems to be
justified to some extent by its success.

We present the results of a numerical evolution of the function 
$F(t,\tau)$, choosing the simplest observable $\cO=\exp(iz)$ and the 
parameter $N_I=0.1$. To do this, both the evolution of the probability 
density $P$ (see Eq.~(\ref{realFPE})) and the evolution of the observable 
(see Eqs.~(\ref{obsevol}),(\ref{obsevol2})) are needed.

$P(x,y;t-\tau)$ is obtained by using the time dependent FPE in the 
Fourier representation; a simple Euler discretization in time with time 
step $10^{-5}$ turns out to be sufficient. This was discussed already in 
some detail in \cite{ass2}.

$\cO(x+iy;\tau)$ is obtained as described in the previous section (Eqs.~ 
(\ref{obsevol}),(\ref{obsevol2})) by using the evolution of $\cO$ under 
$\tilde L$ or equivalently under $L_{y_0}$.  This evolution does not 
depend on either $N_I$ or the cutoff $Y$, since neither $L$ nor $L_{y_0}$ 
depend on those two parameters.

\begin{figure}[t]
\begin{center}
\includegraphics[width=0.95\columnwidth]{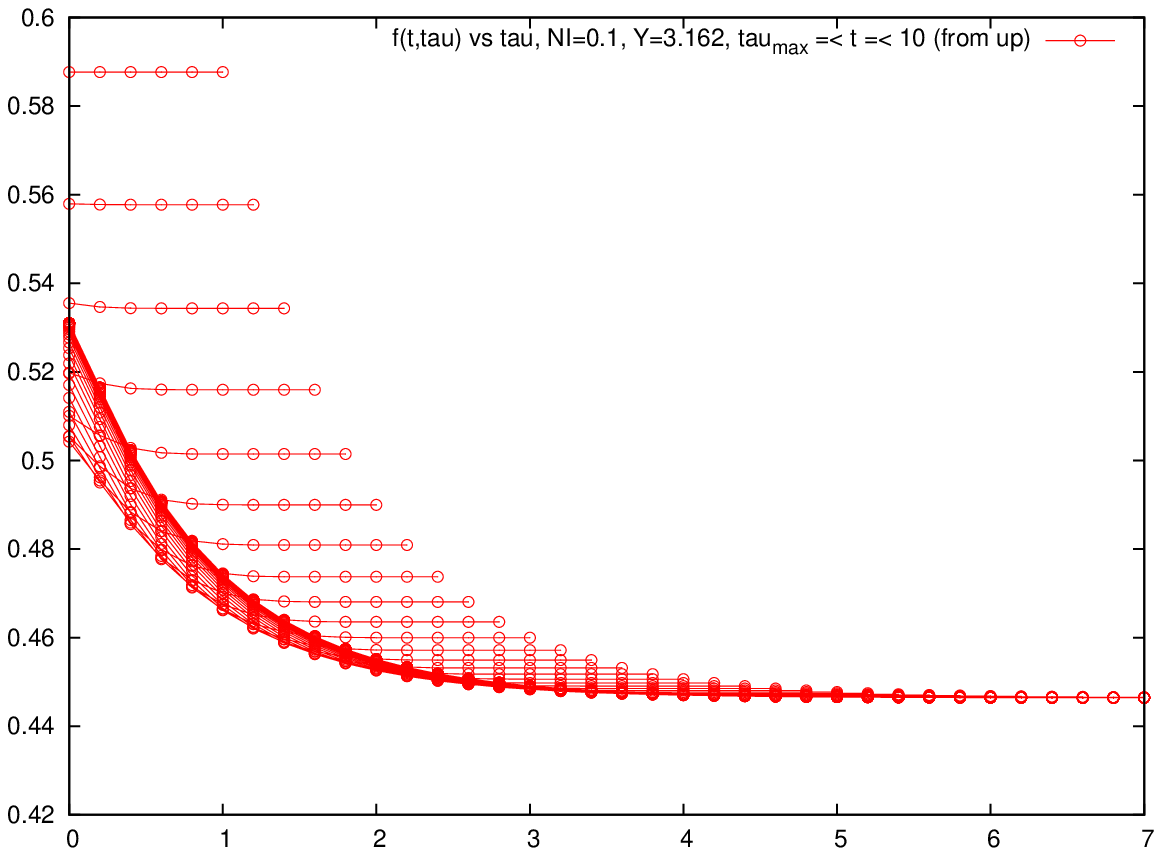}
\caption{$F(t,\tau)$ vs.~$\tau$ for several values of $t$, with
$0<\tau<t$, $N_I=0.1$ and the cutoff $Y=3.162$.}
\label{fttau3.162}
\end{center}
\end{figure}

\begin{figure}[t]
\begin{center}  
\includegraphics[width=0.95\columnwidth,angle=-90]{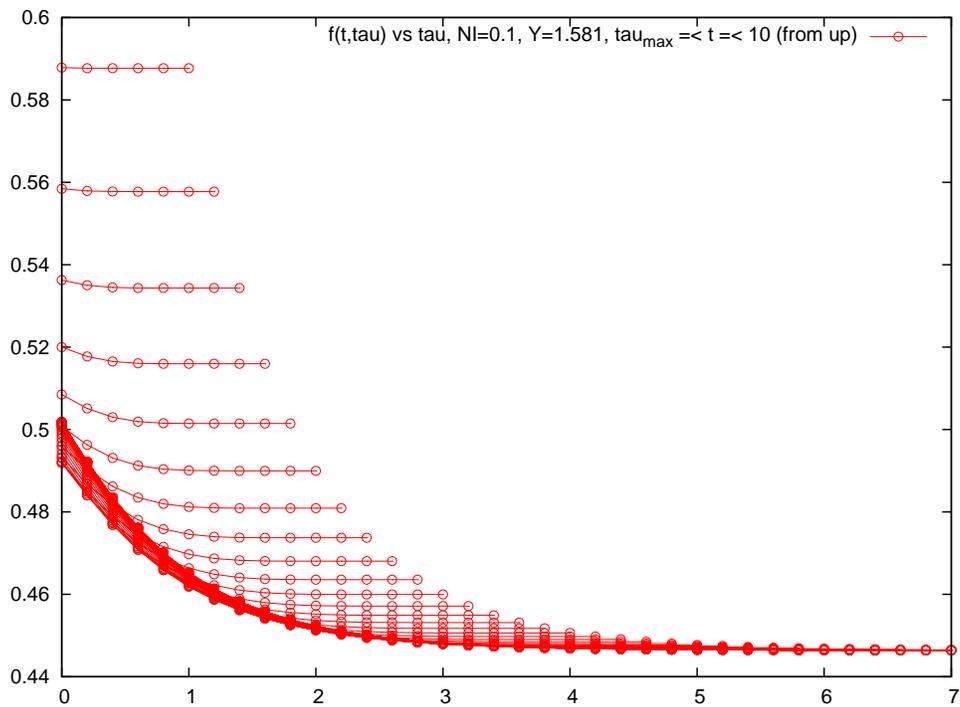}
\caption{As in Fig.~1, with $Y=1.582$.}
\label{fttau1.582}
\end{center}
\end{figure}

\begin{figure}[t]
\begin{center}
\includegraphics[width=0.95\columnwidth]{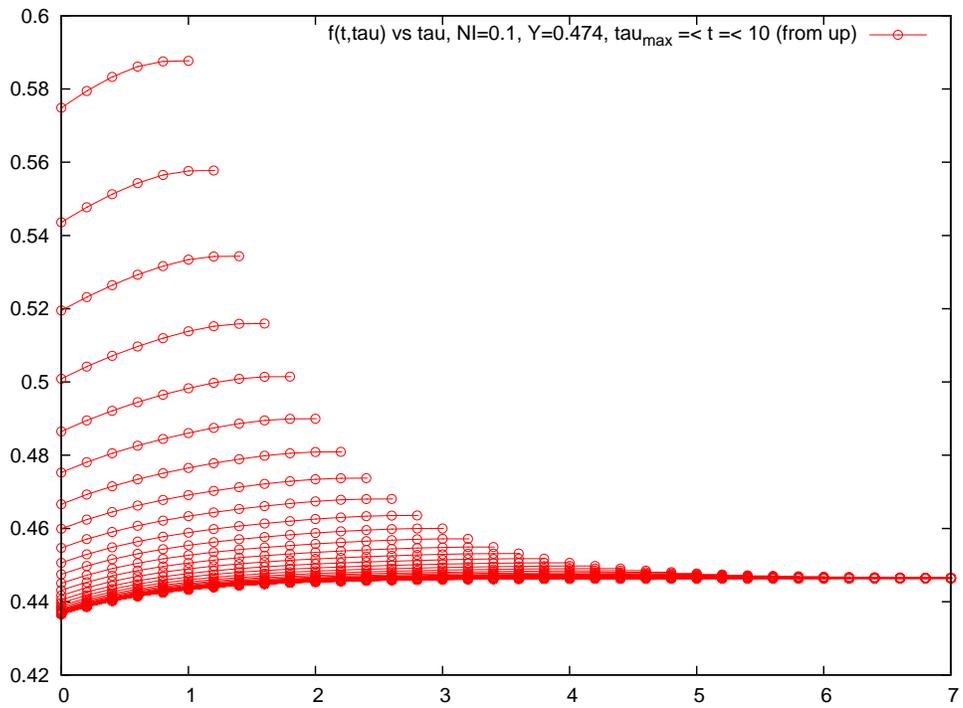}
\caption{As in Fig.~1, with $Y=0.474$}
\label{fttau0.474}
\end{center}
\end{figure}

\begin{figure}[t]
\begin{center}
\includegraphics[width=0.95\columnwidth]{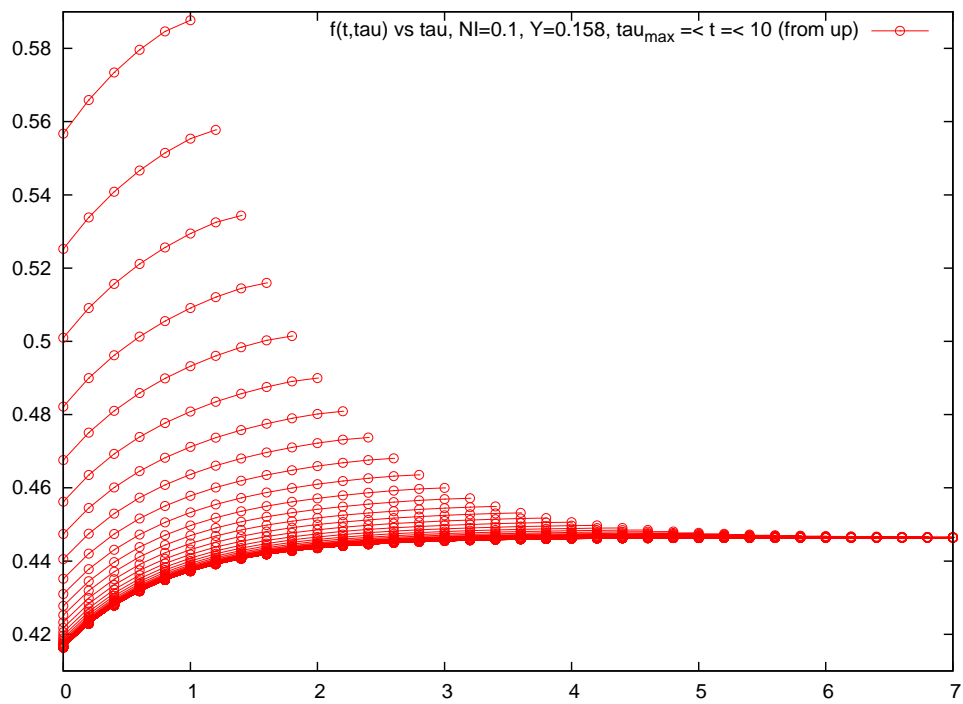}
\caption{As in Fig.~1, with $Y=0.158$.}
\label{fttau0.158}
\end{center}
\end{figure}

$F(t,\tau)$ is then obtained by summing up the products of $\cO(x+iy;t)$ 
and $P(x,y;t-\tau)$. The results are presented in Figs.~\ref{fttau3.162}, 
\ref{fttau1.582}, \ref{fttau0.474}, \ref{fttau0.158}. 

In these plots we show $F(t,\tau)$ as a function of $\tau$, for a number
of $t$ values, ranging from $t=1$ to $t=7$. For every $t$ value, $\tau$
runs from $0$ to $t$. In all cases $N_I=0.1$, while the cutoff $Y$ varies
from $Y=3.162$ in Fig.~1 to $Y=0.158$ in Fig.~3.

The following features can be seen from the figures:

\noindent
(1) In general $F(t,\tau)$ is {\it not} independent of $\tau$,

\noindent
(2) the dependence is always strongest at $\tau=0$,

\noindent
(3) the sign of the $\tau$ derivative changes somewhere between
$Y=0.474$ and $Y=1.582$; there seems to be a `best choice' of cutoff
at which the derivative vanishes.

\begin{figure}[t]
\begin{center}
\includegraphics[width=0.95\columnwidth]{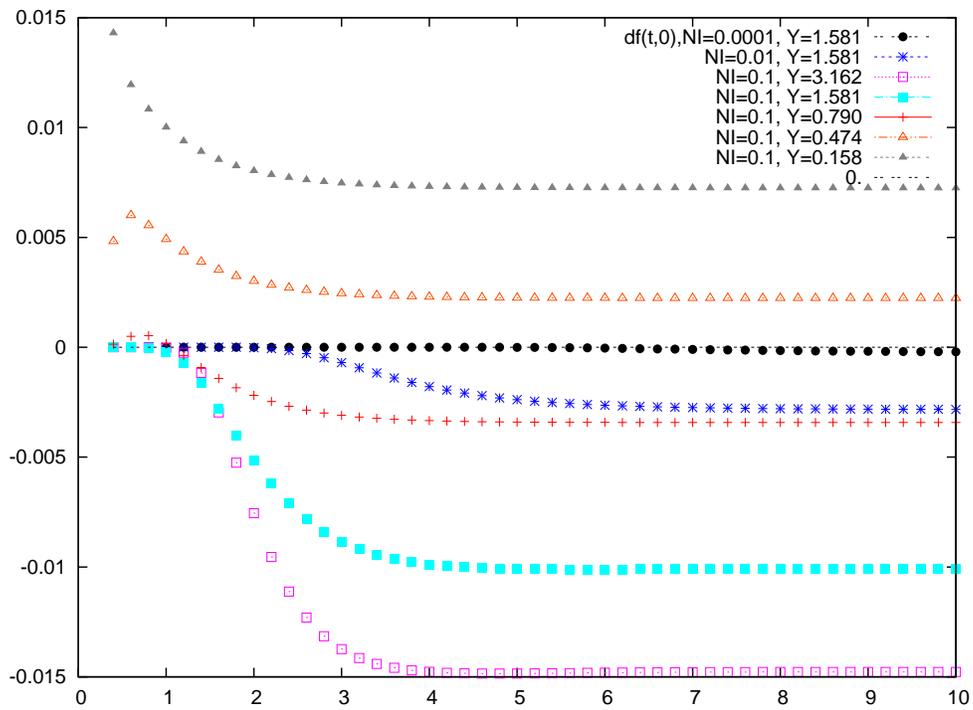}
\caption{$\tau$ derivative of $F(t,\tau)$ as a function of $\tau$
at $t=10$ for several values of the cutoff $Y$ and complex noise 
parameter $N_I$.}
\label{fttauderiv}
\end{center}
\end{figure}

This picture is corroborated by Fig.~\ref{fttauderiv}, which shows 
directly the $\tau$ derivatives obtained as finite difference 
approximations. In this figure we also show different values of $N_I$ and 
it is clearly visible that for very small values of $N_I$ the derivative 
also effectively vanishes. Note that $N_I=0$, which should be preferred 
for a CLE simulation, cannot be used for the FPE computations, because it 
would lead to instabilities (see Section VIB below).

\subsection{Mathematical analysis of the failure}

In this subsection we analyze in more detail the behavior of the time 
evolved observables in order to understand why in general $F(t,\tau)$ is 
not independent of $\tau$. 

We describe the evolution of the observables in some more detail: the 
Langevin operator $\tilde L$ is
\be
\tilde L=\frac{d^2}{dz^2}-a \sin (z-ic)\frac{d}{dz}\,.
\ee
For the observables $e^{ikz}$ we find
\be
\tilde L e^{ikz}=-k^2 e^{ikz}-\frac{a}{2}k\left(e^c e^{i(k+1)z}-
e^{-c}e^{i(k-1)z}\right)\,.
\ee
Choosing now $c=0$ and $a=\beta$, we consider an observable
\be
\cO(z)=\sum_k a_k e^{ikz}\,
\ee
and its time evolution $\cO(z;t)\equiv \sum_k a_k(t) e^{ikz} $ defined by
Eq.~(\ref{obsevol}). This evolution can be expressed in terms of the
coefficients $a_k$ as follows:
\be
\label{modeobsevol}
\partial_t a_k(t)=-k^2 a_k(t)+\frac{\beta}{2}\bigl[-(k-1)a_{k-1}(t)
+(k+1)a_{k+1}(t)\bigr]
\ee
and may be viewed as evolution under $\tilde L$, $L$ or, if we fix $y=0$,
as evolution under $L_0$. The evolution operator $L_0$ in Fourier space is
thus represented by a tridiagonal matrix with elements
\be
\left(\widehat L_0\right)_{kk'}=-k^2\delta_{kk'}+\frac{\beta}{2}
\left[-(k-1)\delta_{k-1,k'}+ (k+1)\delta_{k+1,k'}\right]\,.
\ee

We now establish the following facts:

\noindent
(1) The Langevin operators $L_{y_0}$ generate exponentially bounded
semigroups on the Hilbert space $L^2(dx)$ for any $y_0$. In particular
there are no poles.

\noindent
(2) If the Fourier transform of $\cO$ contains only positive modes, this
will also be true for $\exp(tL_{y_0})\cO$. But typically then all positive
modes will be populated.

\noindent
(3)
\be
\lim_{t\to\infty} e^{tL_{y_0}}\cO= \frac{1}{Z_{y_0}}\int dx\, \cO(x+iy_0)
e^{-S(x+iy_0)}\,
\ee
and the convergence is exponentially fast.

\noindent
(4)
For holomorphic observables $\cO$
\be
\exp(t L)\cO= \exp(tL_{y_0})\cO\,.
\ee
Since the right hand side is independent of $N_I=N_R-1$, so is the left
hand side. This argument does not involve any integration by parts.

\noindent
(5) $\cO(x+iy;t)$ grows for $t>0$ more strongly than any exponential as 
$y\to\-\infty$, invalidating integration by parts except for $N_I=0$.

The proof of (1) follows from a theorem to be found in \cite{daviesbook}
(Theorem 11.4.5). The point is that the drift (first order in derivatives) 
term of $L_{y_0}$ is a so-called Phillips perturbation of the Laplacian:
\be
L_{y_0}=A+B\,,
\ee
with
\be
A=\frac{d^2}{dx^2}\,,\quad B=\beta \sin(x+iy_0)\frac{d}{dx}\,.
\ee
$B$ can be applied to any vector of the form $\exp(t A)\psi$, $t>0$
and we have
\be
\label{phillips}
\int_0^1 dt \Vert B\exp(t A)\Vert<\infty\,.
\ee
These two properties allow to set up a perturbation expansion for
$\exp[t(A+B)]$ and show its convergence. Explicitly
\be
e^{t(A+B)}=e^{tA}+
\sum_{n=1}^\infty
\int_{0\le t_1\le\ldots\le t_n\le t}
e^{t_1A}Be^{(t_2-t_1)A}B\ldots B e^{(t-t_n)A}\,.
\ee
Convergence in norm is not hard to see: by Fourier transformation one sees
that
\be
\Vert \frac{d}{dx} e^{t A}\Vert = \sup_k |ke^{-t k^2}|\le
\frac{1}{\sqrt{2 t e}}\,,
\ee
hence
\be
\Vert Be^{t A}\Vert \le {\rm const}\, \beta\, e^{|y_0|}
\frac{1}{\sqrt{t}}\,.
\ee
From this is it obvious that the bound (\ref{phillips}) holds; since the
integration volume in Eq.~(\ref{phillips}) is $t^n/n!$, the series
converges in norm;

(2) is obvious;

(3) means in particular that the evolution of $\cO$ converges to a 
constant. While it is obvious that all constants are eigenfunctions of 
$L_{y_0}$, we don't have sufficient analytic understanding of the spectra 
of the operators $L_{y_0}$ to prove this convergence. Numerically, 
however, it is seen easily that the evolution converges to the correct 
constant and the convergence is exponentially fast;

(4) is an obvious consequence of analyticity;

(5) is seen by numerically analyzing the growth of the coefficients
$a_k(t)$ for $t>0$:

Using the initial condition $a_1=1$, $a_{k}=0$ for $k\neq 0$ and $\beta=1$ 
as 
before, $a_k(t)$ are the Fourier coefficients of $\exp(tL_0)\cO_1$ with 
$\cO_1(x)=\exp(ix)$. In Fig.~ref{growth} we plot $-\ln (|a_k(t)|)/k$ for 
four different times ($t=0.5,1,2,3$) against $\ln(k)$. As remarked, only 
positive modes get populated; it turns out that the coefficients $a_k(t)$ 
alternate in sign. From this we conclude that $|\cO_1(z;t)|$ grows most 
for large negative $y$ and is maximal for $x=\pm\pi$. Modes were cut off 
at $|k|=50$, but the picture shows for all the times clearly an  
asymptotic linear increase with a slope close to 1, so we conclude
\be
\label{growthcoeff}
a_k(t)\sim K^k (-1)^k k^{-\gamma k}\,,
\ee
with $\gamma$ possibly slightly less than 1 and some constant $K$. Further 
numerical studies show that the behavior of Eq.~(\ref{growthcoeff}) is 
universal: it is independent of the initial condition and $\beta$. For 
comparison in this figure we also show (in black) the quantity $\ln 
(k!)/k+\ln(2)$, which seems to be approached asymptotically by the other 
curves. 

Since Eq.~(\ref{growthcoeff}) obviously implies
\be
\label{growthcoeff2}
|a_k(t)|\ge K^k k^{-k}\,,
\ee
by a simple argument we can conclude that $\cO(x+iy;t)$ grows 
superexponentially in $y$ direction: we put $w=e^z$; then, using only 
positive modes for the initial conditions, $\cO(z;t)$ is given by the 
power series
\be
\cO(z;t)=\sum_{k=0}^\infty a_k(t) w(z)^k\,.
\ee
Cauchy's estimate says that for any $R\ge 0$
\be
|a_k(t)|\le S(R)R^{-k}\,,
\ee
where
\be
S(R)= \sup_{|w|=R}  | \cO(z(w);t) |=
\sup_{x} | \cO(x-i \ln R;t) |\,.
\ee
From this and our numerics we conclude that asymptotically
\be
S(R)\ge (KR)^k k^{-k} \,
\ee
and this holds for any $k$. The optimal value is
\be
k_0=(KR)e^{-1}\,,
\ee
 which leads to the bound
\be
S(e^{-y})=|\cO(\pi+iy)|\ge
\exp\left[{\rm const}\exp(-y)\right]\,.  
\ee
Note that this holds in particular for $y<0$! Since for $N_I>0$ and
$t,\tau>0$ one can at best expect a Gaussian decay of $P(x,y;t)$,
Eq.~(\ref{fttau}) in this case involves an integral of a function that is
not absolutely integrable and hence its value is ambiguous, depending on
the order of integrations. Thus the formal argument for correctness of
the CLE fails.

\section{Falloff of equilibrium measures}
\label{secV}

In this section we study the $t\to\infty$ limit of $P(x,y;t)$, 
i.e.~the equilibrium measure in order to check why and how our general 
criterion Eq.~(\ref{crit}) can fail. As remarked in Section \ref{secIII}, 
the equilibrium condition
\be
L^T P(x,y;\infty)=0
\ee
implies fulfillment of the criterion
\be
E_{\cO}\equiv\int P(x,y;\infty) \tilde L \cO(x+iy;0)dxdy=0\,,
\ee
provided integration by parts on $\cM_c$ without boundary terms at 
imaginary infinity is justified. So the falloff of $P(x,y;\infty)$ is 
crucial for success or failure.

\subsection{U(1) one-link model}

For the U(1) one-link model studied in \cite{Aarts:2008rr,ass2} we are 
able to make rather precise statements about the falloff of the 
equilibrium measure in the $y$ direction. 

The system is symmetric under the reflections $x\mapsto -x$ and $y-c 
\mapsto -(y-c)$. To study the falloff of the equilibrium measure in $y$ we 
again chose $c=0$ and grouped the data obtained by the CLE simulation into 
bins $|y|\in [(n-1/2)\epsilon,(n+1/2)\epsilon)$ with $\epsilon=0.1$. For 
clarity we chose rather large values of $N_I=0.1,0.5,1.0$ and $9.0$. The 
results are shown in Fig.~\ref{u1loghisto} and show clearly a universal 
decay rate 
\be 
P(x,y;\infty)\sim \exp(-2|y|)\,. 
\ee 
This result improves considerably the statement made in \cite{ass2} and 
also explains the difficulties with determining reliably expectation 
values of $\exp(ikz)$ for $|k|\ge 2$ (they are suffering from extremely 
large fluctuations).

In \cite{ass2} we considered the Fourier modes
\be
\widehat P_k(y;t)= \int dx\, e^{ikx} P(x,y;t);
\ee 
formally the expectation values of the exponentials are given by 
\be
\bra e^{ikz}\ket= \int dy\, \widehat P_k(y;t) e^{-ky}\,,
\ee
using the fact that 
\be
\int dy\widehat P_k(y;t)=\int dx dy\, P(x,y;t)=1\,.
\ee
We simplify the notation for $\widehat P_k(y;\infty)$ to $\widehat 
P_k(y)$. By binning in $y$ as above we also produced estimates of 
the modes $\widehat P_k(y)$ for $k=1,2$ and $N_I=1$, shown in 
Fig.~\ref{loghistomodes}. $\widehat P_2$ seems already to be quite noisy, 
but at least the first few kinks visible in the figure for $\widehat P_2$ 
correspond to true sign changes. But what is more important is the 
clearly visible fact that $\widehat P_1$ and $\widehat P_2$ decay 
at least like $\exp(-3|y|)$. This can be confirmed using the 
stationary Fokker-Planck equation (FPE) obeyed by $P(x,y;\infty)$. In 
terms of the Fourier modes the FPE reads (see Eq.~(65) of \cite{ass2}):  
\begin{align}
\label{fpemode}
(N_R k^2- N_I\partial_y^2)\widehat P_k(y)+\,&\frac{\beta}{2} 
\cosh(y)\Big[(k-1)\widehat P_{k-1}(y)
-(k+1)\widehat P_{k+1}(y)\Big]\notag\\
-\,&\frac{\beta}{2}\sinh(y)\partial_y\Big[\widehat P_{k-1}(y)+
\widehat P_{k+1}(y)\Big]=0\,.
\end{align}

Since we are interested in the large $|y|$ asymptotics, we may replace
$\cosh(y)$ and $\sinh(y)$ by $\pm 1/2 \exp(|y|)$. Integrating 
Eq.~(\ref{fpemode}) for $k=0$ from $0$ to $y$ and using evenness in $y$ we 
obtain
\be
N_I {\widehat P}_0'(y)+\frac{\beta}{2} e^{|y|} \widehat P_1(y)=0\,.
\ee
So if $\widehat P_0$ decays like $\exp(-2|y|)$, $\widehat P_1$ will decay 
like $\exp(-3|y|)$. Continuing inductively and assuming exponential 
decay, one obtains easily
\be
\widehat P_k(y)\sim c_k e^{-(|k|+2)|y|}\,.
\ee
Unfortunately Eq.~(\ref{fpemode}) also implies that $c_{k+1}\sim k c_k$, 
which means that one cannot sum up the asymptotic behavior of the 
$\widehat P_k$ to obtain the aymptotics of $P(x,y;\infty)$.

More important is what we learn about the expectation values of 
$\exp(ikz)$, which should be given by
\be
\bra e^{ikz} \ket =\int P(x,y;\infty) e^{ikx-ky} dx dy\,. 
\ee
The integral on the right hand side does not converge absolutely for 
$|k|\ge 2$, hence its value is ambiguous. A well defined result may be 
obtained by first integrating over $x$, but it is not clear if this 
corresponds to the long time average of the complex Langevin process.  
But it seems that the large fluctuations observed in the CLE data 
reflect the fact that the integral is ill defined. 

One can also try to compute expectation values using the binning 
employed above. This corresponds to first integrating over $x$, then 
over $y$. The results, however, agree with those obtained directly by 
the CLE simulation (up to some loss of precision due to the finite 
width of the bins) and not to the exact values. This is of course no 
surprise, as the binning is based on the CLE simulation.  

The conclusion is that the CLE process with complex noise and without 
a field cutoff will in general not produce unambiguous results for the 
expectation values of exponentials $\exp(ikz)$ with higher $|k|$.

\subsection{The model of Guralnik and Pehlevan}

To see if this phenomenon of slow decay of the equilibrium distribution is 
not just a specialty of our U(1) one-link model, we also analyzed the 
equilibrium measure for the simplest polynomial model (called GP model in 
the sequel), studied already by Guralnik and Pehlevan \cite{gp1} and  
discussed briefly in \cite{ass2}.

The model is defined by the action 
\be
\label{GPaction} 
S = -i\beta\left(z+\frac{1}{3}z^3\right)\,;
\ee
since this action is purely imaginary, we have to deform the real axis to 
a path (submanifold) $\cM_r$ as described in Section II such that
$\exp(-S)$ is absolutely integrable over $\cM_r$. A possible choice 
(cf.~\cite{gp1}) is the path $z=x+i\epsilon \sqrt{1+x^2}$ for some small 
positive $\epsilon$.

Since the action produces a stable fixed point at $x=0,\,y=1$, we produced 
histograms representing $P(x,y;\infty)$ by binning $r=\sqrt{x^2+(y-1)^2}$ 
in intervals of length $0.1$. They are shown in Fig.~\ref{gploghisto}. 
Since in this case we expect a power falloff, we use a log-log scale. The 
indications are again that the rate of falloff is the same for different 
values of $N_I>0$, namely roughly like $r^{-1.5}$, whereas for $N_I=0$ we 
find a stronger falloff (we cannot decide at this point whether it is 
still power-like or stronger).

Accepting this observation one concludes that for $N_I>0$ again the 
higher moments $\bra z^k\ket $ of the equilibrium distribution are 
ill-defined, a fact that is reflected by large fluctuations of these 
quantities in the CLE simulations \cite{ass2}.

\section{Testing the criterion}

We now proceed to test the truncated version of our criterion on the two 
toy models introduced; our primary interest is to see whether checking it 
only for a few low moments (modes) is sufficient to identify incorrect 
results.

\label{secVI}

\subsection{U(1) one-link model}

For this model we considered the two cases 
\be
\beta=1,\quad \kappa=0 \, 
\ee
and 
\be
\beta=1,\quad \kappa=0.25,\quad \mu=0.5 \, 
\ee
(which is equivalent to $\beta\approx 1.27,\;\kappa=0$). In both cases we 
chose $N_I=0.1$ which was found in \cite{ass2} to lead to manifestly 
incorrect results for the CLE simulation without cutoff. Using the FPE as 
well as the CLE simulations, we measured the expectation values $\bra 
\exp(iz) \ket$ and $\bra \exp(2iz) \ket$ as well as the corresponding 
quantities $\bra \tilde L \exp(iz) \ket$ and $\bra \tilde L \exp(2iz) 
\ket$. Again we introduced a periodic cutoff $Y$ in imaginary direction 
which stabilizes the FPE solution as well as the CLE expectation values.

In Figs.~\ref{cutoffkappa0} and \ref{cutoffkappa.25} we show $\bra 
\cO_k\ket$ divided by its exact value minus 1 and $E_k$, both for $k=1,2,3 
$. The results indicate the remarkable fact that at a particular value of 
the cutoff not only all the indicators
\be
E_k\equiv\bra \tilde L\exp(ikz) \ket 
\ee
vanish but also 
\be
c_k\equiv\bra \cO_k\ket = \bra \exp(ikz) \ket 
\ee
agree with their exact values (it should be noted that due to the 
symmetry of the system the observables $\exp(-iz)$ and $\exp(-2iz)$ do 
not contain any extra information). Note that $E_2$ has a second zero, 
but at that point $E_1\neq 0$.

So in this case our simple test of the identity (\ref{crit}) for two 
observables is apparently sufficient to identify the correct simulation: 
it has sufficient sensitivity to reject wrong solutions. To make sure that 
at the properly tuned cutoff value the measure $P$ is indeed correct, one 
would in principle have to check all exponentials, again a practical 
impossibility.

In our U(1) one-link model the SDE hierarchy amounts just to the well 
known recursion relation for the Bessel functions $I_k(\beta)$ and it is 
determined by fixing $\bra 1\ket=1$ and $\bra \exp(iz)\ket=c_1$. In a 
CLE simulation $c_1$ will depend on the value of the cutoff. If
\be
c_1\neq\frac{I_1(\beta)}{I_0(\beta)}\,,
\ee
the SDE recursion rapidly runs away to infinity and it is manifest that 
the bound Eq.~(\ref{bound}) cannot hold. So this bound seems to be 
crucial for picking out the right solution of the SDE. On the other hand 
the cutoff models in general obey the bound, but unless the cutoff is 
tuned correctly, they will miss the right value of $c_1$ and fail to obey 
the SDE recursion.

\subsection{Guralnik-Pehlevan model} 

We next apply our test to our other toy model, the cubic model of 
Guralnik and Pehlevan. Since this model has noncompact real and imaginary 
parts, we introduce {\it two} periodic cutoffs: $X$ for the real and $Y$ 
for the imaginary part.

In this model $\tilde L = \partial_z^2+ i\beta (1+z^2)\partial_z$, and the
first few relations read
\bea
E_1&\equiv\bra\tilde L z\ket = i\beta\bra 1+z^2\ket\notag\\
E_2&\equiv\bra\tilde L z^2\ket = 2\bra 1+i\beta z(1+z^2)\ket\notag\\
E_3&\equiv\bra\tilde L z^3\ket = 3\bra 2z+i\beta z^2(1+z^2)\ket\,,
\eea
leading to SD relations between the expectation values of $z^k$.

It is easy to see that the exact results, which can be expressed in terms 
of Airy functions (see \cite{gp1}) and for $\beta=1$ are  $\bra z\ket
\approx 1.1763i$, $\bra z^2\ket =-1$, $\bra z^3\ket\approx -0.1763i$, 
indeed satisfy these relations.

We measured the moments $c_k\equiv\bra z^k\ket$ for $k=1,2,3,4$; this 
allows also to obtain $E_2=\bra\tilde L z^2\ket$ and $E_3=\bra\tilde L 
z^3\ket$; note that $E_1=\bra\tilde L z\ket=i+i\bra z^2\ket=0$ is already 
tested by comparing $\bra z^2\ket$ to its exact values $-1$. In 
Fig.~\ref{GPcutoff} we present the results obtained for $N_I=1$ and a 
fixed cutoff $X=3.17$ in $x$ direction, both by using the FPE and the CLE 
simulation. For this value of $N_I$ it was observed already in \cite{ass2} 
that CLE without cutoff does {\it not} reproduce the correct values. The 
figure, on the other hand, shows that there is a value of the $y$ cutoff 
(near $Y=0.8$) for which the two criteria $E_2=E_3=0$ are fulfilled and 
also the right values for the moments $c_1,c_2,c_3,c_4$ are obtained.

With purely real noise ($N_I=0$) the situation is quite different. For 
this case the FPE simulation is unstable: Fig.~\ref{instability} shows the 
time evolution of the FPE for $N_I=0$; the evolution settles onto 
metastable values very close to the exact ones, but then takes off and 
diverges. For comparison we also show the FPE time evolutions for two 
rather small nonzero values: $N_I=0.01$ and $N_I=0.1$ (all three figures 
are using the cutoffs $X=Y\approx 3.95$). As seen in 
Figs.~\ref{instability}, \ref{stability0.1}, the small imaginary noise 
is 
sufficient to stabilize the evolution, at least for the times considered. 
This seems to conform at least qualitatively to the discussion found in 
Numerical Recipes Ch.~19 \cite{nrc19}. Quantitatively from that 
discussion one would expect that much larger values $N_I$ would be needed 
for stabilization; we are lucky that this does not seem to be the case 
here and in fact with this small nonzero $N_I$ we obtain good convergence 
to the exact result, provided the cutoff is not extremely small.

The CLE simulation, on the other hand, works perfectly for $N_I=0$. We 
have seen already in Section \ref{secIII} that for $N_I=0$ the equilibrium 
distribution is quite well concentrated and shows a very strong falloff. 
In agreement with this, we find that the data are quite insensitive to the 
cutoffs introduced; for $X=3.95$ as before, even a cutoff of $Y=0.8$ is 
sufficient to produce values close to the exact ones and consequently also 
fulfill the criteria $E_1=E_2=E_3=0$ with good precision. These facts can 
be clearly seen in Fig.~\ref{GPcutoff0.0}; in this figure we display for 
comparison the CLE results for $N_I=0$ and the FPE results for $N_I=0.01$ 
(recall that $N_I=0$ does not allow for a viable FPE solution).

Again we found that our simple test seems to have sufficient sensitivity
to select the right simulation.

\section{Conclusions and outlook}
\label{secVII}

In this paper we have pinned down the reasons why the CLE simulations 
sometimes fail to produce correct expectation values and we have developed 
practical tests for correctness. In two toy models we checked the strength 
of the test, by deforming the process through introduction of complex 
noise ($N_I>0$) as well as cutoffs. It turned out that our tests are 
successful in picking out the correct results.

In the context of this paper the introduction of a nonzero $N_I$ plays a
double role:\\
(1) As a means to check the applicablity of the formal proofs for
correctness of the CLE.\\
(2) As a parameter which can be used for tuning and stabilizing the
simulation. As such it is needed in FPE computations, but for true lattice
models, where there is little chance to use the FPE anyway, it is
probably still best to stay with $N_I=0$. The periodic field cutoff   
introduced in the toy models should be seen in a similar way.

One should not, however, expect that this simple device of tuning a cutoff 
will work in general to produce correct results. Other modifications might 
be necessary, but the most promising choice is still to work with the 
unmodified CLE process and purely real noise. 

The main point is that our results demonstrate the `sensitivity' of the 
truncated test criterion, in addition to the `specificity' which holds on 
general grounds.

While we studied here only two very simple models, we believe the reasons
for incorrect results identified here apply much more generally. They 
are:\\
$\bullet$ rapid growth of the Langevin evolved observables in imaginary 
direction,\\
$\bullet$ slow decay of the equilibrium distribution.

The study of the issues discussed in this paper will be continued; both 
the etiology and the diagnostics will by studied in the XY model 
\cite{ajss} and in nonabelian models \cite{ajsss}.

\vspace*{0.3cm}
\noindent
{\bf Acknowledgments}

I.-O.~S.\ thanks the MPI for Physics M\"unchen and Swansea University for 
hospitality. G.~A.\ and F.~A.~J. are supported by STFC.

\newpage

\begin{figure}[t]
\begin{center}
\includegraphics[width=\columnwidth]{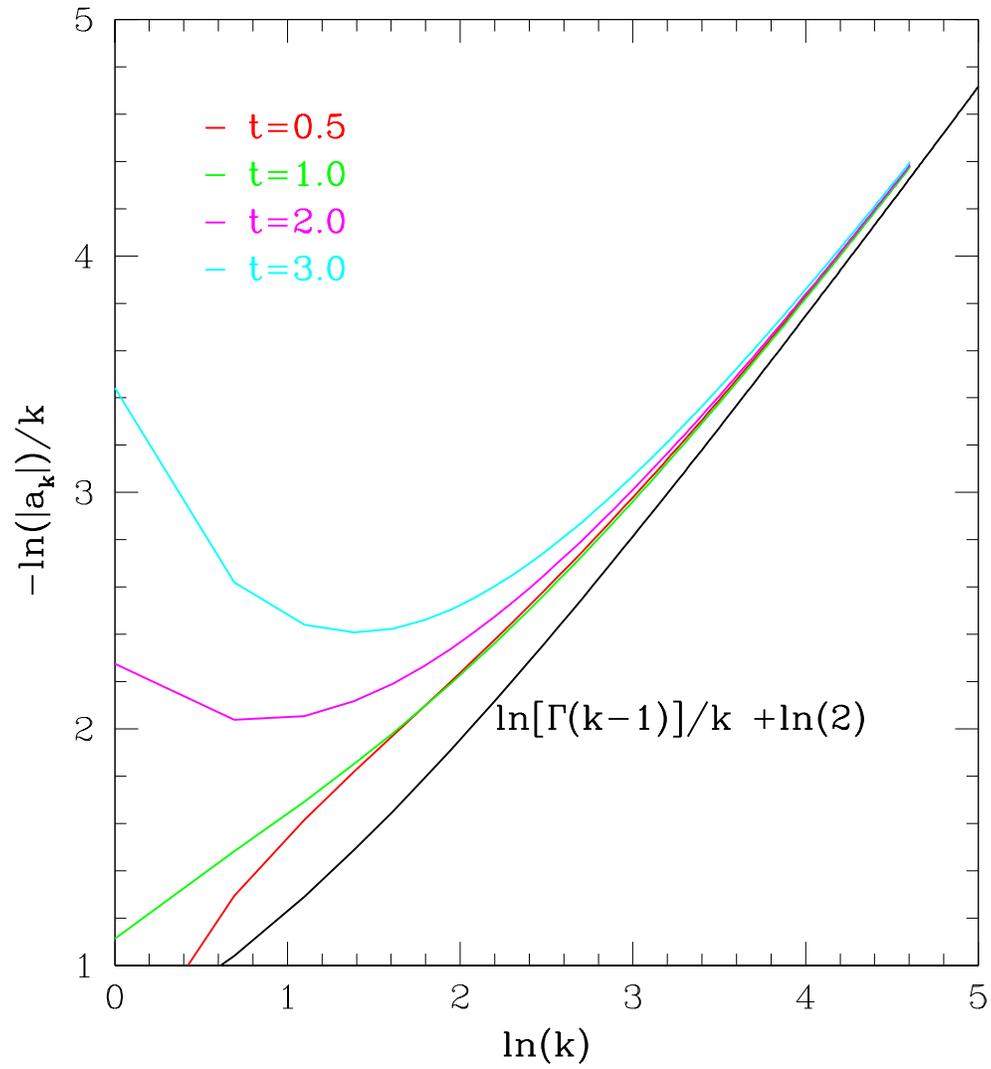}
\caption{Asymptotics of the Fourier coefficients $a_k(t)$.}
\label{growth}
\end{center}
\end{figure}

\begin{figure}[t]
\begin{center}
\includegraphics[width=\columnwidth]{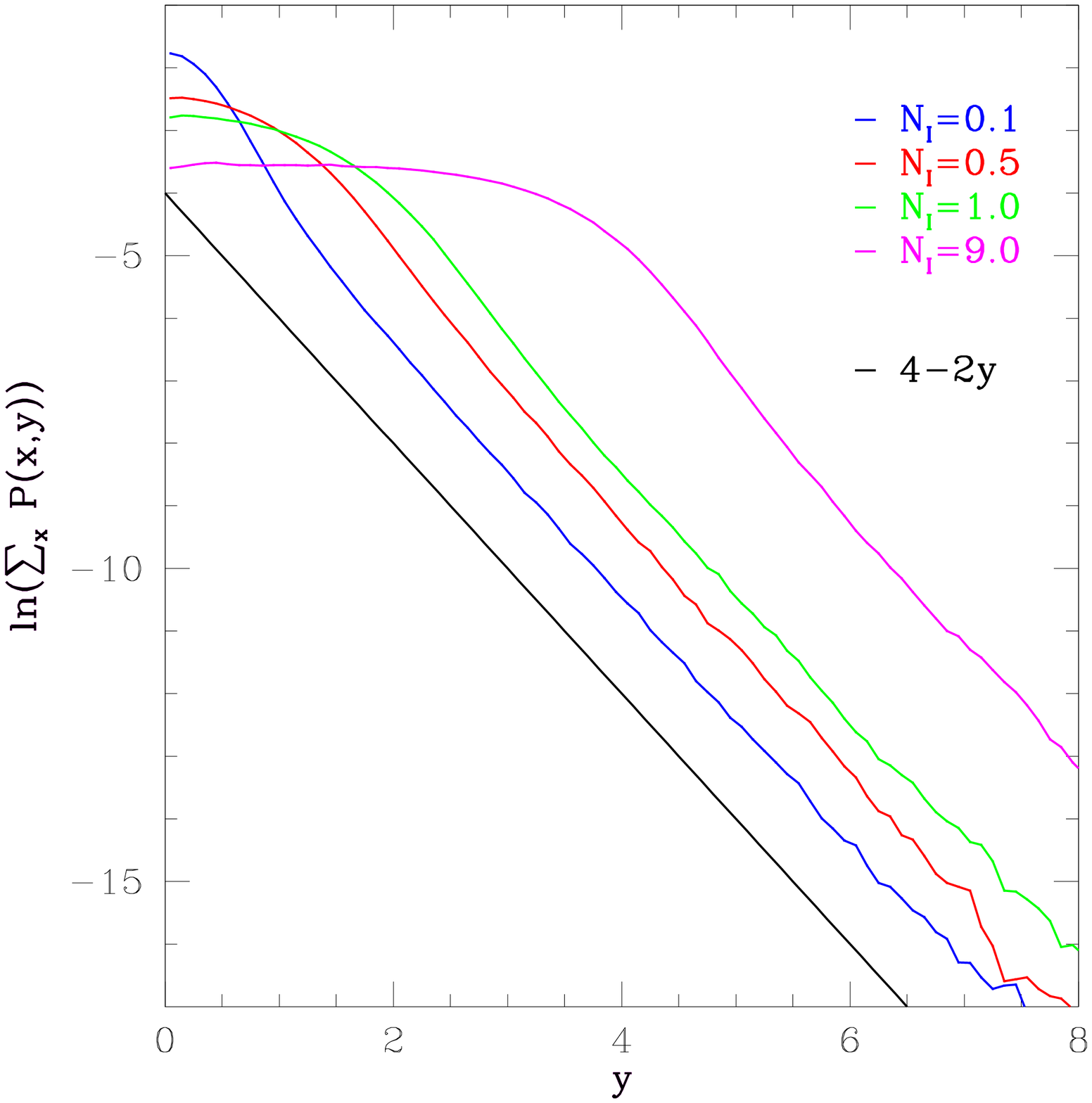}
\caption{Logarithmic histograms of the equilibrium measures for the U(1) 
one-link model.}
\label{u1loghisto}
\end{center}
\end{figure}

\begin{figure}[t]
\begin{center}
\includegraphics[width=\columnwidth]{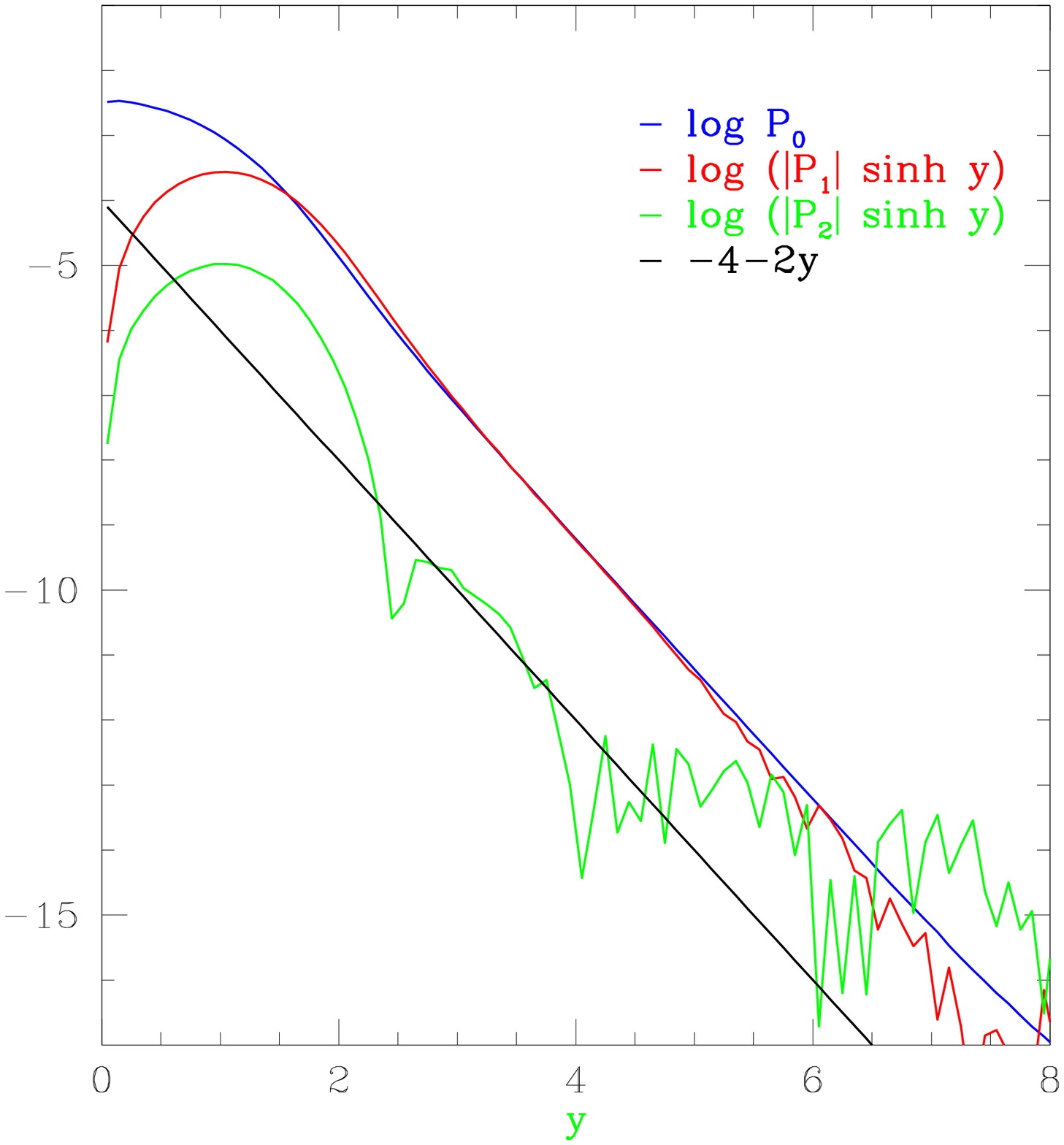}
\caption{Logarithmic histograms of the low modes of the equilibrium
measures for the U(1) one-link model.}
\label{loghistomodes}
\end{center}
\end{figure}

\begin{figure}[t]
\begin{center}
\includegraphics[width=\columnwidth]{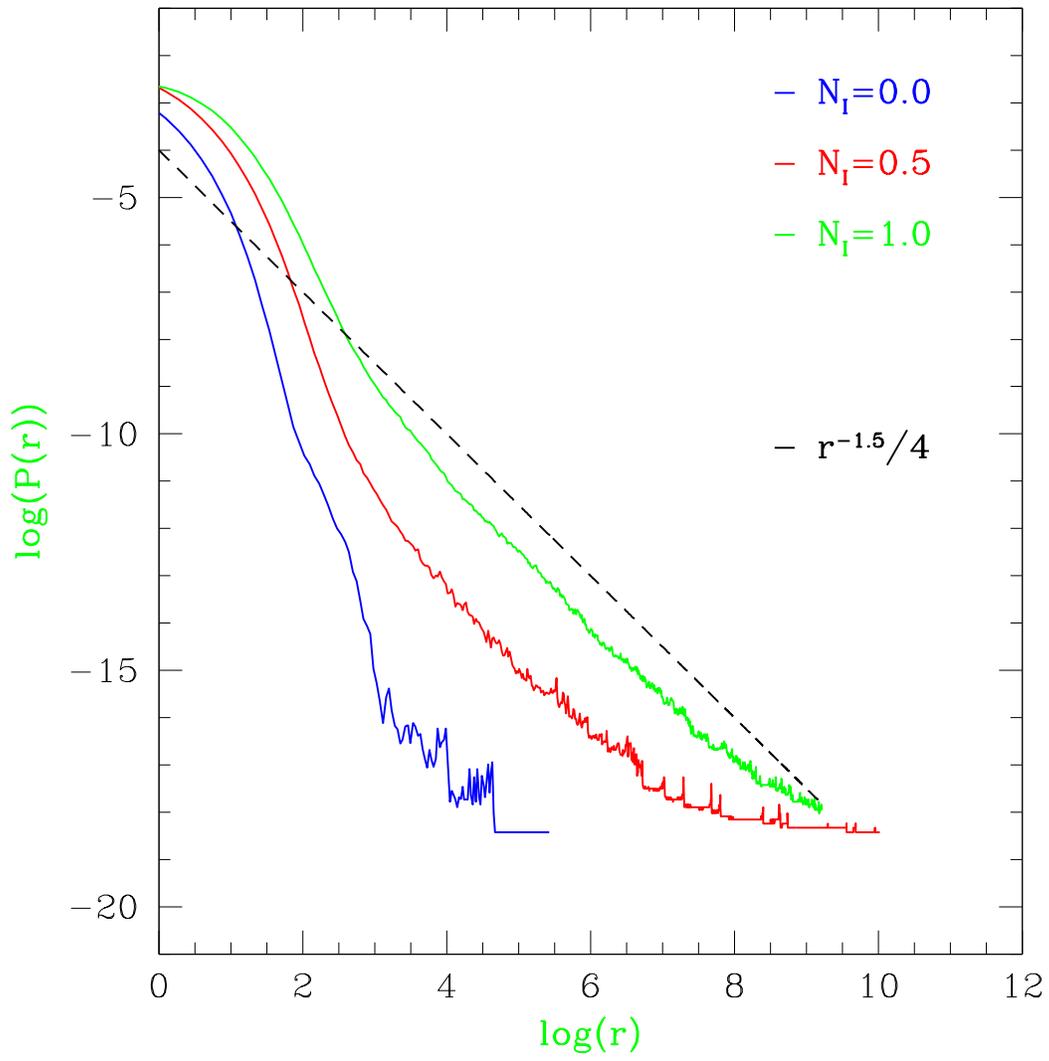}  
\caption{Histograms of the equilibrium measures for the GP model on a
log-log scale.}
\label{gploghisto}   
\end{center}
\end{figure}

\begin{figure}[t]
\begin{center}
\includegraphics[width=0.95\columnwidth]{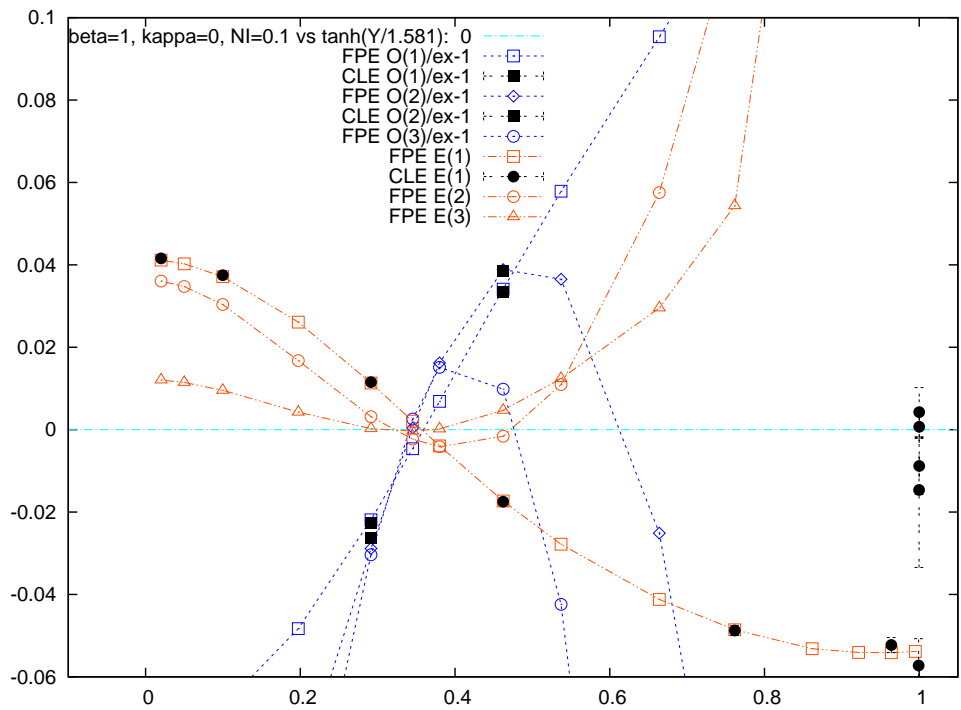}
\caption{Cutoff ($Y$) dependence of various quantities in the
U(1) one-link model, for $\beta=1$, $\kappa=0$, $N_I=0.1$. See main text
for further details.}
\label{cutoffkappa0}
\end{center}
\end{figure}

\begin{figure}[t]
\begin{center}
\includegraphics[width=0.95\columnwidth]{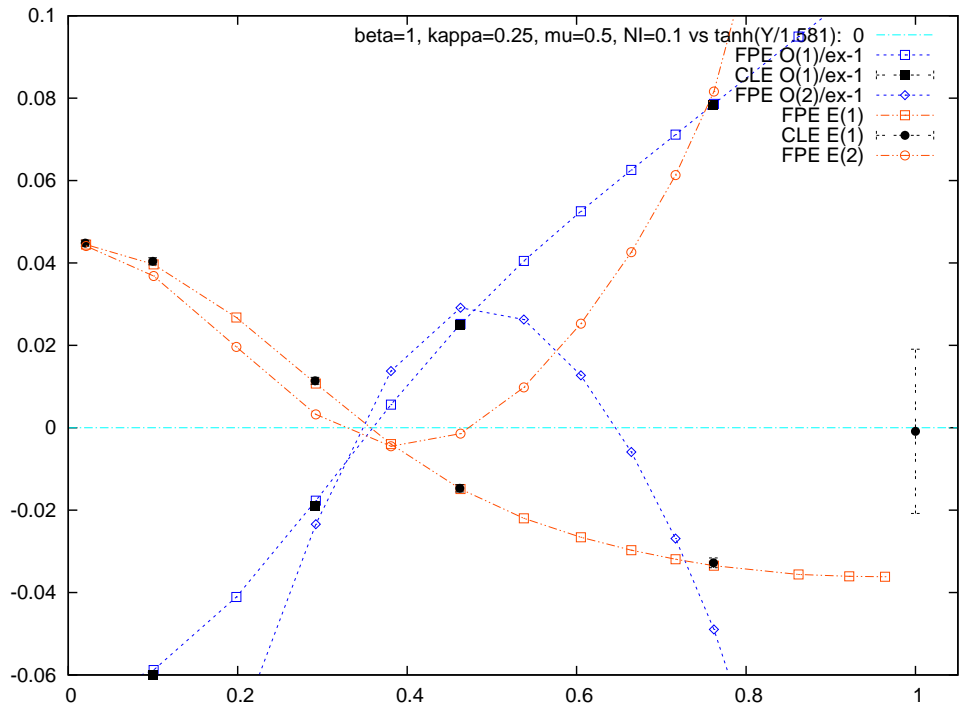}
\caption{As in Fig.~\ref{cutoffkappa0}, for $\beta=1$, $\kappa=0.25$ and 
$\mu=0.5$.}
\label{cutoffkappa.25}
\end{center}  
\end{figure}

\begin{figure}[t]
\begin{center}
\includegraphics[width=0.95\columnwidth]{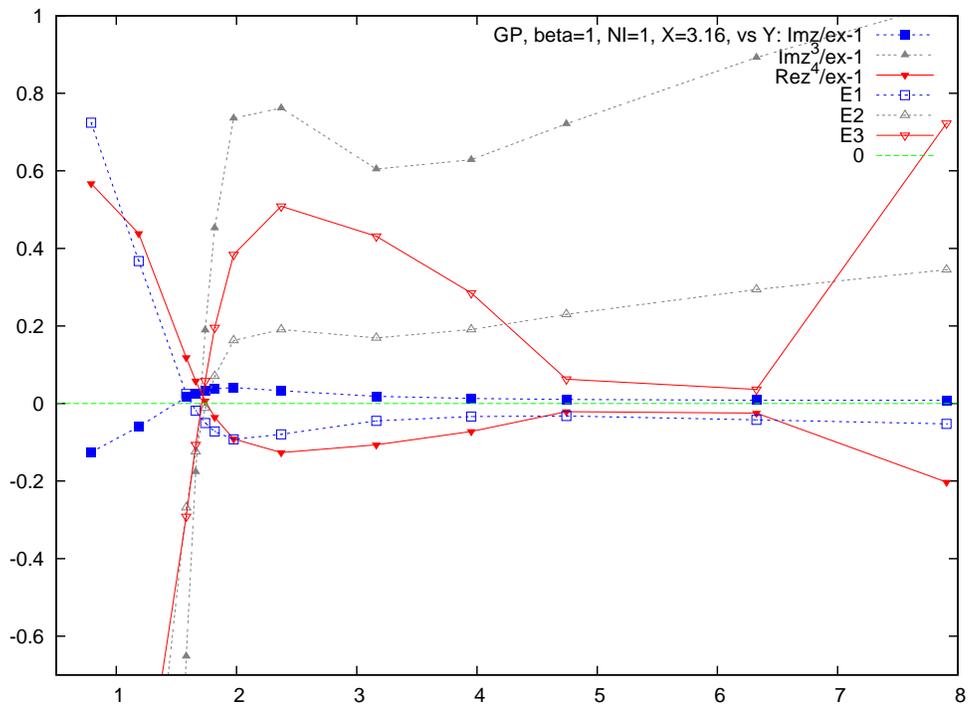}
\caption{GP model: Cutoff dependence for $\beta=1, N_I=1$.}
\label{GPcutoff}
\end{center}
\end{figure}

\begin{figure}[t]
\begin{center}
\includegraphics[width=0.95\columnwidth]{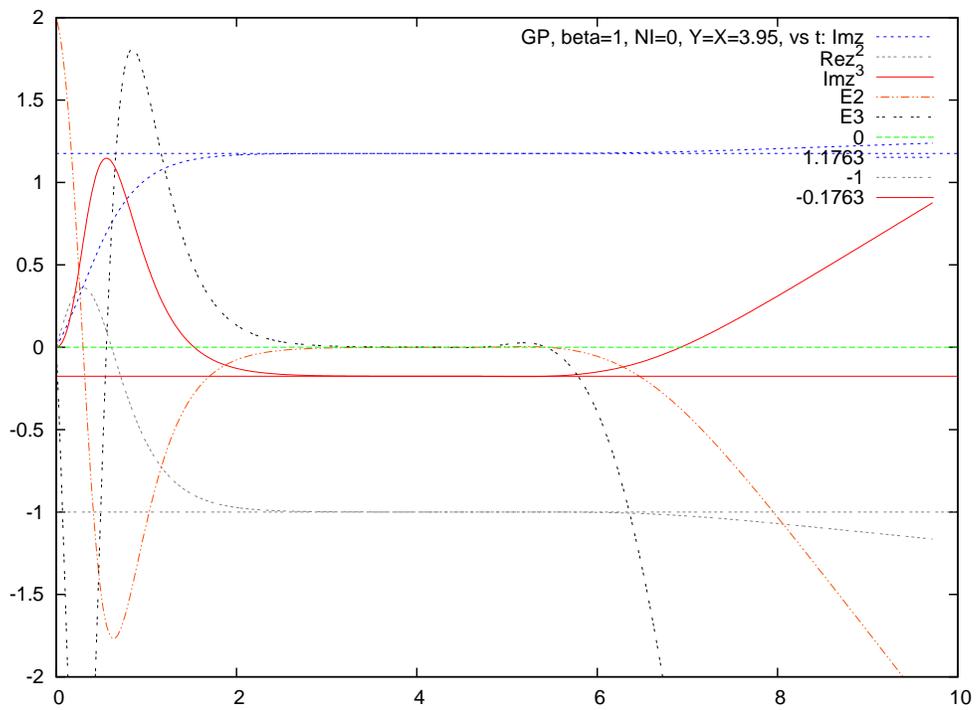}
\caption{FPE evolution in the GP model: $t$ dependence of
various quantities for $N_I=0$, $\beta=1$ and cutoffs $X=Y=3.95$. See
main text for more details.}
\label{instability}
\end{center}
\end{figure}

\begin{figure}[t]
\begin{center}
\includegraphics[width=0.95\columnwidth]{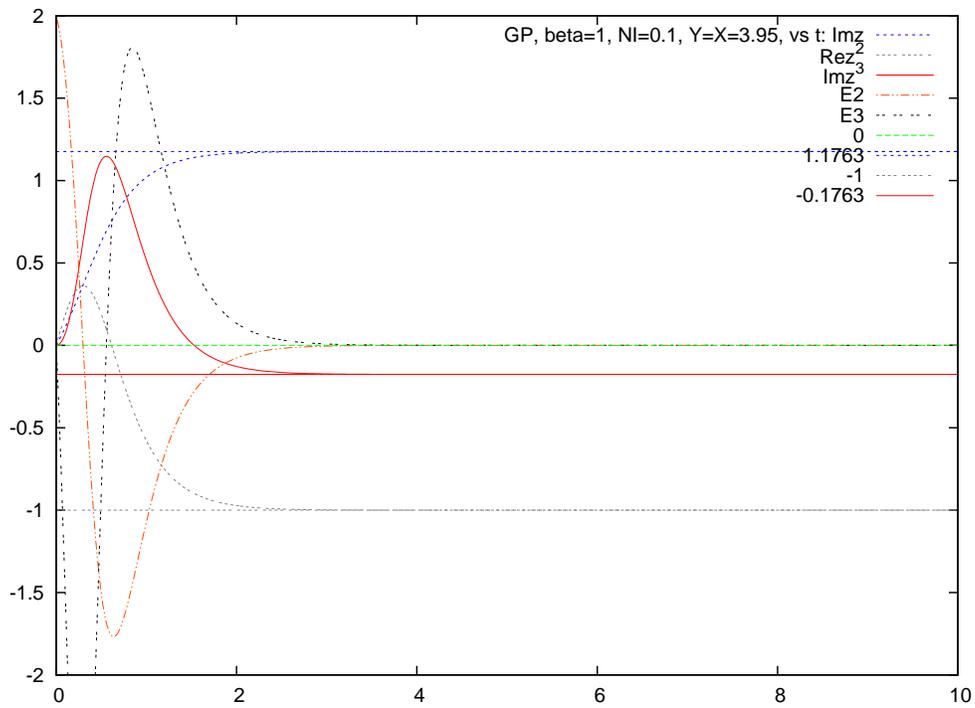}
\caption{As in Fig.~13 for $N_I=0.1$.}
\label{stability0.1}
\end{center}
\end{figure}

\begin{figure}[t]
\begin{center}
\includegraphics[width=0.95\columnwidth]{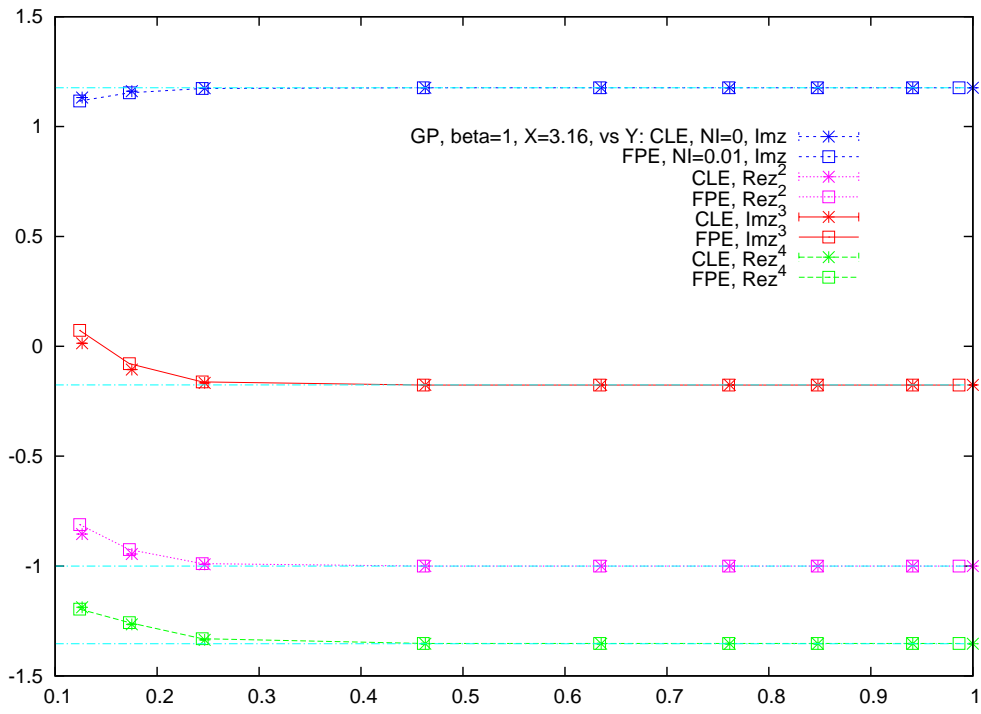}
\caption{GP model: Cutoff ($Y$)dependence for $\beta=1$ and $N_I=0$
(CLE) and $N_I=0.01$ (FPE)}
\label{GPcutoff0.0}
\end{center}
\end{figure}

\end{document}